\documentclass[12pt]{article}

\usepackage{amsmath,amsfonts,amssymb,latexsym,graphicx}
\usepackage{caption}
\usepackage{subcaption}

\def\be{\begin{equation}}
\def\ee{\end{equation}}
\def\bq{\begin{eqnarray}}
\def\eq{\end{eqnarray}}
\def\beq{\begin{eqnarray}}
\def\eeq{\end{eqnarray}}

\def\pa{\partial}

\begin{document}
\title{\textsc{Dynamical synchronization, the horizon problem, and initial conditions for inflation}}
\author{\Large{\textsc{Spiros Cotsakis}\thanks{skot@aegean.gr}}\\
Institute of Gravitation and Cosmology\\ RUDN University,
ul. Miklukho-Maklaya 6, Moscow 117198, Russia\\
Research Laboratory of Geometry,  Dynamical Systems  and Cosmology\\
University of the Aegean, Karlovassi 83200, Samos, Greece}
\date{April 2023}
\maketitle
\begin{abstract}
\noindent We consider the evolution of homogeneous cosmologies towards the future in a dynamical systems formulation. Using a variational equation approach, we show that there is a short period  in which transient solutions between the end of a Mixmaster era and a subsequent Friedmannian state exist. Implications about the generic inhomogeneous evolution towards the future, the recollapse problem, the horizon problem, and the initial conditions required for inflation are briefly discussed.
\end{abstract}
\newpage

\section{Introduction}
\begin{figure}
\includegraphics[width=0.9\textwidth]{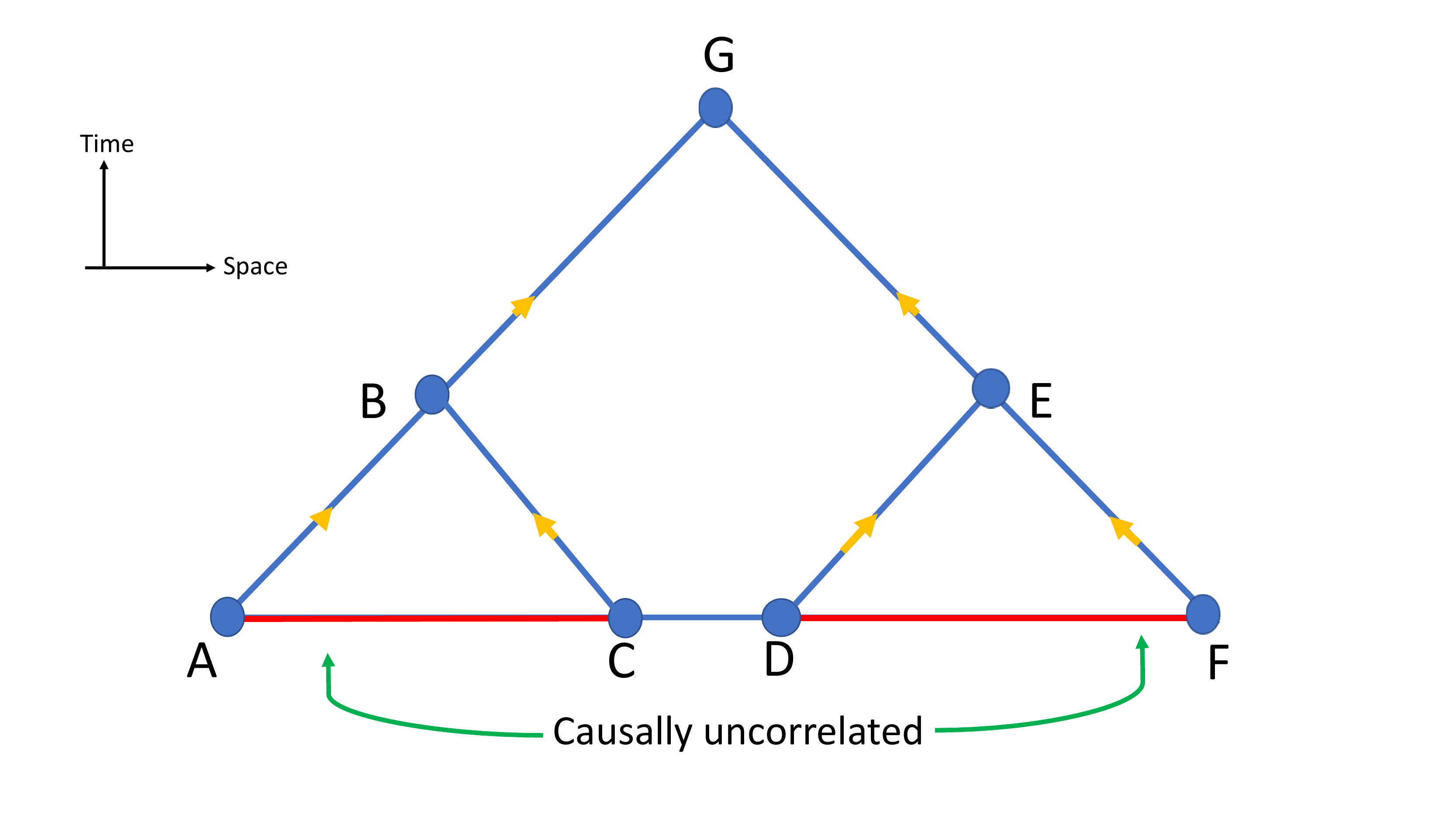}
\caption{According to the horizon problem, rays  such as $CBG$ and $DEG$ transport to G the news that the two causally disjoint regions $AC$ and $DF$ are completely synced although never in causal contact.}
\label{hor1}
\end{figure}

\noindent The observable part of the universe is at any time comprised of a large number of spatial subregions  causally connected to the observer,  not all of which are so connected to each other. This, as is well known, creates the so-called horizon problem,  why any two subregions in the observer's past causal cone, sufficiently separated as to never had been in causal contact throughout their entire history, now appear synchronized in their physically measurable properties (we shall use the word `sync' for synchronization hereafter).  If  causally disjoint regions were so delicately brought to sameness very early \cite{weinberg1}, pp. 525-6, \cite{mtw}, p. 815, \cite{di-pee}, p. 506., then the universe was  in some very special synchronized state  initially, and the question arises as to why this was so.
The basic situation associated with the horizon problem is described in Fig. \ref{hor1}, where we see a generic snapshot  of an `observer' at $G$ overlooking two spatial points $B,E$ in opposite directions, and the two causally disconnected regions $\mathcal{B}=AC,\mathcal{E}=DF$ placed on the spatial hypersurface $\Sigma$.

Inflationary expansion, as is well known, either through a phase transition or more generally, allows for initially `unsynchronized' regions  to become homogeneous very early hence explaining the  subtle differences in the measurements of  observables of two such regions  \cite{weinberg2}, Sec. 4.1B, \cite{li90}, p. 54, \cite{muk}, Sec. 5.1. This explanation of the horizon problem is  the result of the causal mechanism of pushing the big bang hypersurface sufficiently earlier, so that the past light cones of the disjoint regions manage to form a non-empty intersection hence allowing them to  `homogenize' (cf. e.g., \cite{emm}, Sec. 9.7.2,  \cite{pen-road}, Sec. 28.3). However, even if one accepts inflation as a solution to the horizon problem, there is still the issue of initial conditions for inflation, the universe must be sufficiently uniform over some large scale before inflation, for it to subsequently inflate. In particular, for large field inflation one does not have to assume homogeneity over a Hubble domain because inhomogeneities redshift \cite{bran}, but for small field inflation there is a problem of initial conditions \cite{gold}. Why did the field have a large averaged value suitable for inflation over many domains at the end of the pre-inflationary era?

This question leads to another one: What is the relation between causality and synchronization at the end of the  pre-inflationary period? During inflation, two  regions needed \emph{only} to be able  to communicate causally with each other at some specific time in their past history to homogenize and be considered as  synced.

Here we introduce \emph{dynamical synchronization} of spatial domains as a transient or temporal analogue of the phase transition used in inflationary models. We show that although causality-induced homogenization may be a sufficient condition for sync, it is by no means a necessary one. In Refs. \cite{ba20,sync1}, a sync mechanism was introduced as a way to drive the universe to simpler states in the past direction on approach to the initial Belinski-Khalatnikov-Lifshitz  (`BKL') singularity \cite{bkl2}.
In this paper, however, we focus our attention to the \emph{future} evolution of the universe  as it exits an early BKL chaotic phase containing different Mixmaster domains in an inhomogeneous background as conjectured in Ref. \cite{bkl3}. In that case, due to an earlier formation of horizons such regions are causally disconnected. However,  we show that they are able to gradually  `absorb' each other and for some finite time interval of their future evolution resemble a single Friedmannian domain and so proceed in symphony (for an analogous effect observed in the context of biological oscillators, see \cite{winfree,stro2,stro1}).

The plan of this paper is as follows. In the next Section, we establish notation, and show that the evolution equations we use admit the dihedral group as their symmetry group.  In Section 3, we develop a variational equation approach, and using it we show that there is a finite time interval in which an anisotropic Mixmaster domain evolves to become near-Friedmannian in the future direction. In the last Section, we present arguments that suggest  that the same may happen in the generic inhomogeneous case as the universe exits  the BKL phase and enters a subsequent  Friedmannian stage, and comment on the  possible implications of this model for the recollapse problem, the horizon problem, and the issue of initial conditions for inflation.

\section{Dihedral symmetry}
The main result  of this Section is that the equations that describe the evolution of separate Mixmaster domains are invariant under the action of   the dihedral group $D_3$,   that is the symmetry group of the equilateral triangle, which consists of plane rotations through $2\pi/3$ angles followed by (the so rotated) $x$-axis reflections. The main purpose of this section is to describe the path that leads to this result and to establish the notation that will be used subsequently.

Our treatment uses the orthonormal frame formalism for the Einstein equations in which one obtains evolution and constraint equations relative to an orthonormal frame $(e_0,e_a), a=1,2,3$, with $e_0$ normal to the group orbits, cf. \cite{we}, Section 1.4. For the anisotropic Bianchi models of class A, this formalism leads to a dynamical system  in further `expansion-normalized' variables. If $t$ denotes proper time and $H$ the Hubble scalar, one introduces the $\tau$-time by $dt/d\tau=1/H$, and finds  equations of the form $\dot{X}=f(X)$, with constraint $g(X)=0$, where the prime denotes differentiation with respect to the $\tau$-time, $f$ is a polynomial vector field in $\mathbb{R}^n$, and $g(X)$ is  a smooth function on $\mathbb{R}^n$  (with $n$ relatively small).

More specifically, the Bianchi cosmologies can be described by a spacetime whose metric admits a 3-dimensional isometry group $G_3$ acting simply transitively on the spacelike homogeneous hypersurfaces of the spacetime. The  Lie algebra of Killing vector fields (with basis $\xi_a$, and structure constants satisfying $[\xi_a,\xi_b]=C^c_{ab}\,\xi_c$) is the one associated to the symmetry group $G_3$. The orthonormal frame spatial vectors commute with the $\xi$'s, $[e_a,\xi_b]=0$, thus making the orthonormal frame $(\pa/\pa t,e_a)$ `group-invariant'. Using the symmetric object $\eta^{mn}$ and vector $a_m$, both with constant entries and satisfying $\eta^{mn} a_m=0$ in our case, the commutation  functions $\gamma^c_{ab}(t)$ of the resulting group-invariant, orthonormal frame decompose into the standard form, $\gamma^c_{ab}=\epsilon_{abn}\eta^{mn}+a_a\delta^m_b-a_b\delta^m_a$, with $\epsilon_{abc}$ being the alternating symbol with $\epsilon_{123}=+1$. Then ones arrive at the standard classification of the  Bianchi cosmologies into ten different group types using the eigenvalues of the matrix $\eta^{mn}$ (cf. \cite{we}, p. 37).

In the resulting  orthonormal frame approach, the metric components $g_{00}=-1, g_{0a}=0$, and the spatial ones are given by the identity matrix, $g_{ab}=\delta_{ab}$, so that the basic gravitational variables are the non-zero commutation  functions $\gamma^c_{ab}(t)$, of which now the only remaining ones are: the Hubble scalar $H$, the diagonal components of the shear tensor $\sigma_{ab}$ which, being traceless, can be parametrized by only  two independent functions $\sigma_+, \sigma_-$, and the three diagonal components of the matrix $\eta_{ab}=\textrm{diag} (n_1,n_2,n_3)$. Setting $X(\tau)=(N_1,N_2, N_3, \Sigma_+,\Sigma_-)$  where the $N$'s are the normalized spatial scale factors $n_a/H$, while the $\Sigma$'s are the normalized $\sigma$'s, $\sigma_\pm/H$, one obtains the following the evolution equations \cite{wh,we}:
\bq
N'_1&=& (q-4\Sigma_+)N_1,\label{n1}\\
N'_2 &=& (q+2\Sigma_+ + 2\sqrt{3}\Sigma_-)N_2, \label{n2}\\
N'_3 &=& (q+2\Sigma_+ - 2\sqrt{3}\Sigma_-)N_3,\label{n3} \\
\Sigma'_+ &=& -(2 - q)\Sigma_+ -3S_+, \label{+}\\
\Sigma'_-&=& -(2 - q)\Sigma_- - 3S_-,\label{-}
\eq
with the constraint,
\be\label{constraint}
\Sigma_+^2+\Sigma_-^2 +\frac{3}{4}(N_1^2+N_2^2+N_3^2-2(N_1N_2+N_2N_3+N_3N_1))=1,
\ee
where,
\bq
q&=&2(\Sigma_+^2+\Sigma_-^2),\label{q}\\
S_+&=&\frac{1}{2}\left((N_2-N_3)^2-N_1(2N_1-N_2-N_3)\right),\label{splus}\\
S_-&=&\frac{\sqrt{3}}{2}(N_3-N_2)(N_1-N_2-N_3)\label{sminus}.
\eq
In this Section, we prove   that the Wainwright-Hsu system (\ref{n1})-(\ref{sminus}) is $\Gamma$-equivariant, where $\Gamma$ is  the $\Sigma_+$-reflection subgroup of the dihedral group $D_3$. This then implies that the system is $D_3$-equivariant.

We begin with some definitions. Let $\Gamma$ be a compact Lie group, and we consider its action on $\mathbb{R}^n$. We say that the vector field $f$ is $\Gamma$-\emph{equivariant} with respect to the action of $\Gamma$ provided that for each $\gamma\in\Gamma$ and $X\in\mathbb{R}^n$, we have,
\be \label{equiv}
f(\gamma X) =\gamma f(X).
\ee
In this case, for any solution $X$ of the $\Gamma$-equivariant system $\dot{X}=f(X)$, it follows that $\gamma X$ is also a solution.

Now let $D_3$ be the dihedral group of order 6 acting on the $(x,y)$-plane $\mathbb{R}^2$. This is the symmetry group of an equilateral triangle with vertices $(1,2,3)$ on the plane, where for concreteness, we assume that the triangle is positioned such that the vertex 1 is somewhere on the negative $x$-axis, and 2, 3 are on the 4th and 1st quadrants respectively. 

Then $D_3$ consists of the three (counterclockwise) rotations $r_0,r_1,r_2$ about its center of angles $0,2\pi/3,4\pi/3$ respectively, and the three reflections $s,u,t$ with $s$ being the reflection about the $x$-axis, that is $s(123)=132$, $u$ the reflection about the axis that forms an angle of $\pi/3$ with the  $x$-axis, that is $u(123)=213$, and $t$ is the reflection about the the axis that forms an angle of $2\pi/3$ with the $x$-axis, that is $t(123)=321$. Then $r_0=I$ is the identity, while the remaining five elements of $D_3$ can be simply described with the help of the rotation and reflection matrices of the plane.

We now identify the vertices of the equilateral triangle with $N_1,N_2,N_3$, and the $x,y$ axes with $\Sigma_+,\Sigma_-$, respectively. If we set $\Gamma_s=\left\{I,s\right\}$, that is the subgroup of  $D_3$ that consists of the identity and the $s$ reflection of $D_3$, it is not difficult to show  that  the system  (\ref{n1})-(\ref{-}) is $\Gamma_s$-equivariant. That is if $(N,\Sigma)$ is any given solution, then $s(N,\Sigma)$ is also a solution. This follows by direct calculation (or even by a simple inspection), since when the $s$ element is applied and we get $s(N_1N_2N_3)=N_1N_3N_2$, we also have,
\be
s\left(
  \begin{array}{c}
    \Sigma_+ \\
    \Sigma_- \\
  \end{array}
\right)
=
\left(
  \begin{array}{cc}
    1 & 0 \\
    0 & -1 \\
  \end{array}
\right) \left(
  \begin{array}{c}
    \Sigma_+ \\
    \Sigma_- \\
  \end{array}
\right),
\ee
so that the whole system (\ref{n1})-(\ref{-})  is unchanged under $s$ (the constraint (\ref{constraint}) is also invariant in that case). (We note that under other subgroups, $\Gamma_t=\left\{I,t\right\}$,  $\Gamma_u$, etc, the system  is \emph{not} equivariant.)\footnote{The fact that the $s$ reflection is a symmetry of the equations was first noted in \cite{wh}, cf. their first unnumbered equation in their p. 1416. However, the whole statement in that reference after their Eq. 2.29 is misleading, because the previous stated symmetry in the last unnumbered equation of p. 1415 as representing a `rotation through $2\pi/3$ rad in the $\Sigma_+ - \Sigma_-$ plane' is erroneous. As presented in that reference, the first transformation that appears in the last line in p. 1415 is the element $r_1$,  while the second one written there is the element $r_2$ in our notation (in which case one gets a full rotation around the circle, that is the identity element).}

The property  that the system (\ref{n1})-(\ref{-}), (\ref{constraint}) is $\Gamma_s$-equivariant is important because of the following reason. From the multiplication table of $D_3$, it follows that $r_1\circ u=s=r_2\circ t$, and therefore the system is in fact $D_3$-equivariant, in the sense that it is invariant under $2\pi/3$- and $4\pi/3$-rotations followed by the corresponding $x$-axis (that is $u$, or $t$) reflection ($s$-reflection corresponds to the element $r_0\circ s$ of $D_3$).

$D_3$-equivariance is in turn important because the $D_3$-symmetry of rotations and reflections acting on an `initial' $(N,\Sigma)$ solution as a base, is necessary to sustain the BKL sequence of epochs and eras as one progresses to the $t=0$ singularity in the homogeneous case as developed in \cite{bkl2}. This fact is perhaps not clearly emphasized in the literature, for it is hard to see how the BKL oscillations could be sustained without $D_3$-equivariance. This symmetry becomes even more important for the BKL conjecture in the inhomogeneous case, where the homogeneous BKL behaviour is conjectured to exist all the way to the generic singularity \emph{for each Mixmaster domain}, as discussed in \cite{bkl3,uvwe}. $D_3$-symmetry is then valid on any Mixmaster subregion in inhomogeneous spacetime with  solutions  in distinct subdomains differing only in their  spatially-dependent phases (while preserving the $D_3$ symmetry).

\section{Transient behaviour}
We now consider, besides the solution $X=(N,\Sigma)$, another solution $Y=(M,\Pi)$ satisfying the system (\ref{n1})-(\ref{-}), (\ref{constraint}), with $p=2(\Pi_+^2+\Pi_-^2)$ in the place of $q$ in (\ref{q}),  the spatial curvatures $Q$'s like the $S$'s in  (\ref{splus}), (\ref{sminus}) but with the $M$'s in the corresponding places of the $N$'s, and the constraint identical to (\ref{constraint}) but with the $(M,\Pi)$'s in the places of the $(N,\Sigma)$'s.

For $\tau$ in some open interval $J$, and assuming that the solution $Y$ satisfies the initial condition $Y(\tau_0)=Y_0$, we introduce \emph{the synchronization (or variational) function},
\be \omega=X-Y,\quad \textrm{on}\,\, J,
\ee
and define  \emph{the variational equation along the solution} $Y=(M,\Pi)$,
\be\label{var}
\omega'=D_X f(Y(\tau))\, \omega=0,
\ee
where $D_X f(Y)$ denotes the Jacobian of the  vector field $f(X)$ given by the right-hand-sides of (\ref{n1})-(\ref{-}) evaluated at the solution $Y=(M,\Pi)$. This is a linear equation, with time-dependent coefficients in general.

The basic result  associated with the Eq. (\ref{var}) is that provided the initial condition $\omega(\tau_0)=\omega_0$ is small (that is for $X_0$ near $Y_0$), the function $\omega +Y$ is a good approximation to the solution $X=(N,\Sigma)$ on some \emph{compact} interval $J_0\subset J$ containing $\tau_0$ (for a proof of this result, see Ref. \cite{hs}, pp. 299-302), we note that the \emph{length} of $J_0$ can be anything provided $J_0$ is compact.) That is, if $\phi_\tau (X)$ is the flow of the system  (\ref{n1})-(\ref{-}), and for $\xi$ small we consider the spatial derivative of the flow $\partial \phi_\tau (\tau,Y_0)/\partial X=\omega(\tau,\xi)$, then as $\xi$ tends to zero, the function $Y(\tau)+\omega(\tau,\xi)$ becomes a better and better approximation to $Y(\tau,\xi)$. The former is usually a  better choice than the latter, because $\omega(\tau,\xi)$ is linear in $\xi$. The approximation is uniform in $\tau$ on the compact interval $J_0$, and depends on a smooth variation of the initial condition, $Y_0\rightarrow Y_0+\xi$ (hence the name `variational').

Below we shall be interested  in the behaviour of solutions  of the variational equation for two special choices of the particular solution $Y=(M,\Pi)$, both being equilibrium solutions of the system (\ref{n1})-(\ref{-}), namely:
\begin{itemize}
\item \textbf{EQ-1}: \emph{Friedmann-Lema\^{i}tre point} $\mathcal{F}$,
\be
\Sigma_+=\Sigma_-=0,\quad N_1=N_2=N_3=0,
\ee
\item \textbf{EQ-2}: \emph{Kasner circle} $\mathcal{K}$,
\be
\Sigma_+^2+\Sigma_-^2=1,\quad N_1=N_2=N_3=0,\quad\Sigma_+, \Sigma_{-}: \textrm{constants}.
\ee
\end{itemize}
According to the previous theory, for initial conditions $X_0$ near \textbf{EQ-1} or \textbf{EQ-2} (that is for $\omega_0$ small), the corresponding solutions $X$ of the system (\ref{n1})-(\ref{-}) are well approximated by
those solutions $\omega$ of the variational equation (\ref{var}) for which the Jacobian is evaluated at the equilibrium solutions \textbf{EQ-1} and \textbf{EQ-2} respectively\footnote{We note here the following important fact. For any equilibrium solution $\bar{Y}$ of the system (\ref{n1})-(\ref{-}), the Jacobian $D_X f(\bar{Y})$ is a constant matrix. This matrix, depending on the nature of the equilibrium $\bar{Y}$, may or may not have some eigenvalues on the imaginary axis. Of course, when the equilibrium $\bar{Y}$ is non-hyperbolic, the stability of of the solutions of the system (\ref{n1})-(\ref{-}) near such an equilibrium cannot be described by studying the linearized equation,  and other methods are needed to study the behaviour of the solutions of the system (\ref{n1})-(\ref{-}). However, the importance of the variational equation in this case is that it does describe the \emph{transient or `observable'} behaviour of the solutions of (\ref{n1})-(\ref{-}), that is their behaviour on a \emph{compact} time interval (rather that their long-term behaviour ($\tau\rightarrow \pm \infty$) as implied in stability studies).}.

Let us consider a homogeneous and anisotropic domain that evolves according to the system (\ref{n1})-(\ref{-}).

\begin{figure}
\centering
\includegraphics[width=0.7\textwidth]{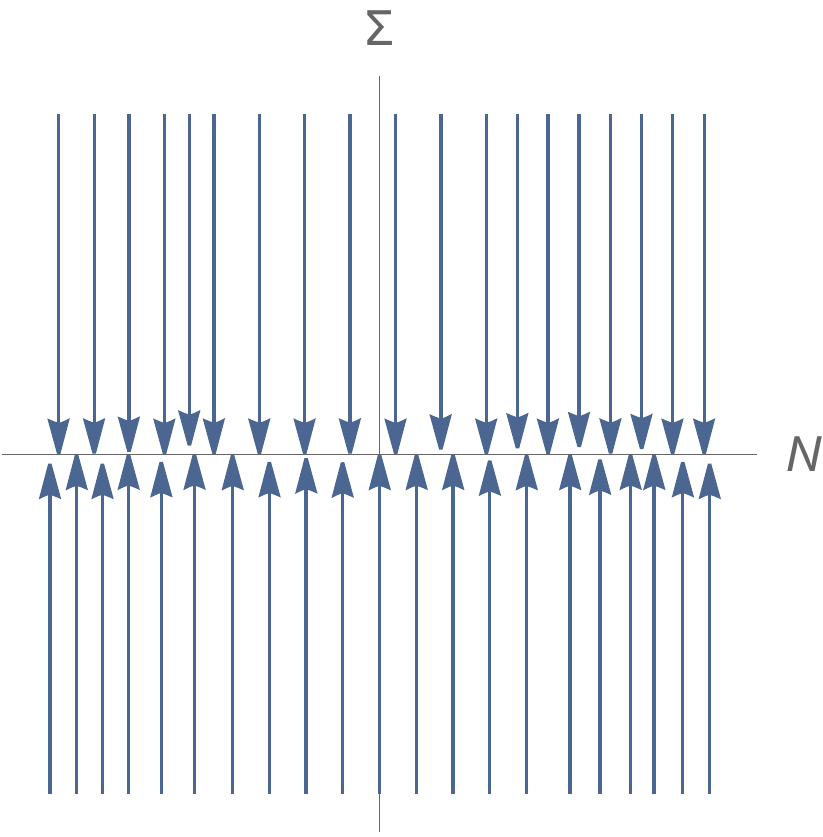}
\caption{Phase portrait giving the behaviour of the orbits near the equilibrium \textbf{EQ-1}.}
\label{eq1}
\end{figure}

At the  equilibrium \textbf{EQ-1}, the Jacobian $D_X f(Y)$ has eigenvalues $0,0,0,-2,-2$ and corresponding generalized eigenvectors the standard basis of $\mathbb{R}^5$. This means that in the $(N,\Sigma)$ phase space, thought as a plane, every point on the $N$ axis is an equilibrium, and all phase points are stably attracted to the $N$ axis along orbits representing parallel lines to the (vertical) $\Sigma$-axis  as in Fig. \ref{eq1}.

Therefore on the  finite time interval $J_0$, all solutions of the system (\ref{n1})-(\ref{-}) in this case are well approximated by Friedmann universes, the exact types of which depend on the specific point of the $N$ axis on which the orbit lands. All such Friedmann universes have been classified, cf. e.g., \cite{we}, p. 129.

To obtain the behaviour of the solutions at \textbf{EQ-2}, we introduce the Kasner exponents $p_i,i=1,2,3$, with $\sum p_i=\sum p_i^2=1$, where in standard notation in terms of the polar angle $\psi$, we have: $p_1=(1-2\cos\psi)/3, p_{2,3}=(1+\cos\psi\pm\sqrt{3}\sin\psi)/3$, with  $\Sigma_+=\cos\psi,\Sigma_-=\sin\psi$.

Then the eigenvalues of the Jacobian $D_X f(\textbf{EQ-2})$ are $6p_1,6p_2,6p_3,0,4$, with corresponding generalized eigenvectors (no relation to the $e$'s of the orthonormal frame previously),   $e_1,e_2,e_3,(0,0,0,-\tan\psi,1),(0,0,0,\cot\psi,1)$, where $e_i, i=1,2,3,$ are the first three vectors of the standard basis of $\mathbb{R}^5$. The  $0$- and $4$-eigenspaces are in the $(\Sigma_+,\Sigma_-)$-plane along the directions defined by the two last eigenvectors, while the  three eigenspaces corresponding to the remaining three eigenvalues are along the $N_1,N_2,N_3$ directions respectively.

The properties of the solutions near \textbf{EQ-2} during their transient passage on the time interval $J_0$, are classified according to  various subsets of the Kasner circle. There are
two cases of such sets on the Kasner circle:
\begin{enumerate}
\item The special case of the so-called Taub points, where two of the Kasner exponents are zero, namely, the $p_i$'s are $(1,0,0), (0,1,0), (0,0,1)$ respectively. In this case the eigenvalues become: $0,0,0,1,4$, and so due to the unstable directions the system  is repelled from  the Kasner circle in the future direction.
\item In all other cases, all of the $p_i$'s are nonzero, with two of them being positive and one negative. This means that the eigenvalues are: $0,-,+,+,4$, that is there is always a zero eigenvalue corresponding to the eigendirection $(0,0,0,-\tan\psi,1)$, one negative eigenvalue depending of which of the $p_i$'s is negative corresponding to one of the eigendirections $N_1,N_2,N_3$, two positive eigenvalues corresponding to the remaining two $N_i$-directions,  and the eigenvalue $4$ along the eigendirection  $(0,0,0,\cot\psi,1)$.
\end{enumerate}
Because of the presence of zero, negative and positive eigenvalues, the system is thus generally unstable in the future direction. In particular,  we can obtain a four-dimensional unstable set for all solutions of the system (\ref{n1})-(\ref{-}) which have the $N_{i_{0}}$ that corresponds to the zero eigenvalue $p_{i_{0}}$ equal to zero. The nature of this invariant set depends on the Bianchi type, and, in particular, orbits with two of the  $N_i$'s zero exist in that set. Along such orbits (with three zero eigenvalues) the system will move away from the Kasner circle and towards the Friedmann state in the future direction.

\begin{figure}
\centering
\includegraphics[width=0.7\textwidth]{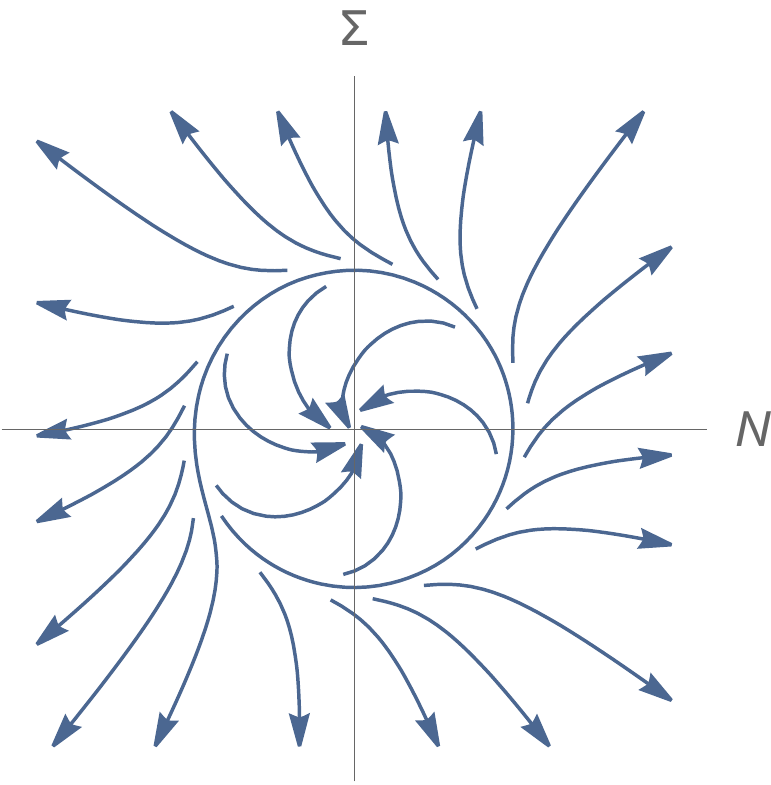}
\caption{Symbolic phase portrait depicting the behaviour of the orbits during the transient period from last BKL era (Kasner circle) to the Friedmann state at the origin.}
\label{trans}
\end{figure}

Therefore the behaviour of the system resembles the one depicted in the symbolic phase portrait of  Fig. \ref{trans}:  the system moves away from the Kasner circle \textbf{EQ-2} in the future because of its instability, and  an exit from the phase of the BKL oscillations  is realized in the future direction where the system is attracted by the equilibrium \textbf{EQ-1}.

This completes the description of the dynamics of the system during the transient period predicted by the variational equation (\ref{var}).

\section{Discussion}
There are various possible implications of these results that we discuss here. One is related to the recollapse problem, another to the horizon problem, and a third one to the impossibility of small field inflation.

Our results show that in the future direction at the end of the BKL period, as a Mixmaster domain evolves off one of the arcs of the Kasner circle of \textbf{EQ-2}, will subsequently be attracted by the Friedmann solution \textbf{EQ-1} and remain close to it for a finite period of time (given by the `observable' time interval we denoted earlier by $J_0$). This describes the mechanism of transient synchronization in a rigorous way (in the future direction and for some finite period of time). 

This result brings some new light to the known property of the Bianchi IX universes in the vacuum case that there are no such universes that expand \emph{forever} and must recollapse, cf. \cite{lw89a}, Theorem 1. This result also provides an alternative mechanism (other than inflation) for the non-premature recollapse  of anisotropic universes until they become isotropic \emph{by transient synchronization} without the need of an inflationary stage. Since it is known that such universes cannot recollapse until they are isotropized by inflation cf. \cite{ba88}, in a sense, our result increases the probability for inflation to occur because it makes the existence of short-lived anisotropic universes that recollapse before inflation less likely. Our result is also to be compared with the long-term behaviour of the Bianchi IX model in the case with a positive cosmological constant, cf. \cite{wa83}, although a non-zero $\lambda$ is not considered here however.

Once one has a Friedmann domain,  what may happen in the next phase of evolution (including the `recollapse' problem discussed above, or  studies of small fluctuations), is not completely addressed either by inflation or by the standard approach to cosmological dynamical systems. This is because one has to take into account \emph{dispersive} aspects present in the structure of the Friedmann equations, and the novel properties  of the universe that are implied in this sense (for more on this, the reader is invited to study Ref. \cite{cot23}).

One may in fact  speculate that in an inhomogeneous setting at the end of the last BKL era, each one of the  different Mixmaster regions involved in the BKL inhomogeneous evolution as they exit the BKL stage in the future direction will, by a similar process, be attracted  by and become a `subdomain' in the attracting Friedmann universe. Hence, the Friedmann domain will contain many such regions which will be causally disjoint because of the formation of horizons during the earlier BKL stage (cf. e.g., \cite{chitre}). The corresponding solutions in other domains will evolve with  different \emph{phases} on the Kasner circle compared to  the specific Mixmaster domain  discussed above, however, one expects that such differences will be erased during the transient period of Friedmannian evolution.

We end up with a transient period from around the end of the BKL stage to the state described by a  Friedmannian domain containing causally disjoined subdomains all synchronized to each other and proceeding in a uniform fashion. As we have already discussed, the length of the transient interval $J_0$  cannot be taken as a stability interval, and so we expect the time duration of  transient  synchronization where the system gets close to Friedmann to be generally  short. This provides an alternative approach to the horizon problem.

On a different front, transient synchronization as described here may have an important consequence for early time - in particular - small field inflation. All the various Mixmaster subdomains that eventually connect to a Friedmannian domain at the end of the transient period, have arisen at the end of the BKL stage as chaotically oscillating regions in the earlier BKL phase in inhomogeneous spacetime. Each one of these Mixmaster subdomains has therefore well-known stochastic properties as described in \cite{klkss}, and consequently, any function defined on the resulting Friedmann domain at the end of the transient period will have its values \emph{randomly} distributed on the set of these Mixmaster subdomains.

If one imagines a scalar field adjoined to the resulting Friedmannian domain, its \emph{initial}, i.e.,  pre-inflationary,  values would thus be expected to be likewise  chaotically distributed, pretty much as dictated by  the Mixmaster subdomains at the end of the last BKL era. Therefore the most natural  value of the scalar field potential energy \emph{averaged over all Mixmaster subdomains}  in the inhomogeneous setup considered at the end of the BKL stage, could naturally be a very large one, of the order of $M^4_{\textrm{Planck}}$, for  classically treated domains of typical minimum size of the order of $M^{-1}_{\textrm{Planck}}$. If true, this implies  that no small field problem of the sort discussed in \cite{bran,gold} could actually be possible at the end of the transient synchronization period. Hence,  inhomogeneous initial conditions natural for inflation are probably expected to occur due to transient synchronization.

\section*{Acknowledgments}
Thanks are due to two anonymous referees for their important comments and suggestions. I am especially  grateful to Robert Brandenberger for his helpful remarks and useful correspondence, and also for providing Ref. \cite{bran}. I thank Jose Mimoso for his warm  hospitality at the Institute of Astrophysics, University of Lisbon,  and John Miritzis for many useful  discussions. This research  was funded by RUDN university,  scientific project number FSSF-2023-0003.


\begin{thebibliography}{99}
\bibitem{weinberg1}S. W. Weinberg, \emph{Gravitation and cosmology} (John Wiley, 1972)
\bibitem{mtw}C. W. Misner, K. S. Thorne, J. A. Wheeler, \emph{Gravitation} (Freeman, New York, 1973)
\bibitem{di-pee}R. H. Dicke and P. J. E. Peebles, \emph{The big-bang cosmology-enigmas and nostrums}, In: General Relativity, An Einstein centenary survey, S. W. Hawking and W. Israel, Eds. (CUP, Cambridge, 1979)

\bibitem{weinberg2}S. W. Weinberg, \emph{Cosmology} (OUP, 2007)
\bibitem{li90}A. Linde, \emph{Particle physics and inflationary cosmology}  (CRC press, 1990)
\bibitem{muk}V. Mukhanov, \emph{Physical foundations of cosmology} (CUP, Cambridge, 2005)
\bibitem{emm}G. F. R. Ellis, R. Maartens, and M. A. H. MacCallum, \emph{Relativistic cosmology} (CUP, 2012)
\bibitem{pen-road}R. Penrose, \emph{The Road to Reality }(Bodley Head, 2004)
\bibitem{bran}R.~Brandenberger,
Int. J. Mod. Phys. D \textbf{26}, no.01, 1740002 (2016)
[arXiv:1601.01918 [hep-th]].
\bibitem{gold}D. S. Goldwirth and T. Piran, Phys. Rep. 214 (1992) 223

\bibitem{ba20}J. D. Barrow, Phys. Rev. D102 (2020) 024017; arXiv: gr-qc/2006.01562
\bibitem{sync1}S. Cotsakis,  Phil. Trans. R. Soc. A380 (2022) 20210189; arXiv: 2010.00298

\bibitem{bkl2}V. A. Belinski, I. M. Khalatnikov, and E. M. Lifshitz, Adv. Phys. 19, 525 (1970)
\bibitem{bkl3}V. A. Belinski, I. M. Khalatnikov, and E. M. Lifshitz,  Adv. Phys. 31, 639 (1982)


\bibitem{winfree}A. T. Winfree,  J. Theor. Biol. 16 (1967) 15
\bibitem{stro2}R. E. Mirollo and S. Strogatz, SIAM J. App. Math. 50 (1990) 1645
\bibitem{stro1}S. Strogatz, \emph{Sync, the emerging science of spontaneous order} (Penguin, 2003)

\bibitem{we}J. Wainwright and G. F. R. Ellis, \emph{Dynamical systems in cosmology} (CUP, Cambridge, 1997)
\bibitem{wh}J. Wainwright and L. Hsu, Class. Quantum Gravity 6, 1409 (1989)

\bibitem{uvwe} C. Uggla, H. van Elst, J. Wainwright and G.F.R. Ellis, Phys.Rev.D 68 (2003) 103502; arXiv: gr-qc/0304002
\bibitem{hs}M. W. Hirsch and S. Smale, \emph{Differential equations, dynamical systems and linear algebra} (Academic Press, 1974)
\bibitem{lw89a}X-f. Lin and R. M. Wald, Phys. Rev. D40 (1989) 3280
\bibitem{ba88}J. D. Barrow, Nucl. Phys. B296 (1988) 697
\bibitem{wa83}R. M. Wald, Phys. Rev. D28 (1983) 2118

    \bibitem{cot23}S. Cotsakis, \emph{Dispersive Firedmann universes and synchronization}, to appear to Gen. Rel. Grav.; arXiv:2208.07892v3
\bibitem{chitre}D. M. Chitre, \emph{Investigation of vanishing of a horizon for Bianchi type IX (the Mixmaster universe), } PhD Thesis, Technical Report No. 72-125, Univ. of Maryland (Faculty Publications Collection, 1972)
\bibitem{klkss}I. M. Khalatnikov, E. M. Lihshitz, K. M. Khanin, L. N. Shchur, and Ya. G. Sinai, J. Stat. Phys. 38 (1985) 97
\end{thebibliography}
\end{document}